\documentstyle[12pt,aps,prl,psfig]{revtex}
\tighten
\begin{document}
\draft 

\title{The Dynamics of Supercooled Silica: Acoustic modes and Boson peak}

\author{J\"urgen Horbach, Walter Kob and Kurt Binder}

\address{Institut f\"ur Physik, Johannes Gutenberg-Universit\"at,
Staudinger Weg 7, D-55099 Mainz, Germany}

\date{\today}

\maketitle

\begin{abstract}
Using molecular dynamics computer simulations we investigate the
dynamics of supercooled silica in the frequency range 0.5-20~THz and the
wave-vector range 0.13-1.1\AA$^{-1}$. We find that for small
wave-vectors the dispersion relations are in very good agreement with
the ones found in experiments and that the frequency at which the
boson-peak is observed shows a maximum at around 0.39\AA$^{-1}$.
\end{abstract}


\section{Introduction}
In the last few years a significant effort was undertaken to understand
the nature of the so-called boson-peak, a prominent dynamical feature
at around 1~THz which is observed in strong
glassformers~\cite{boson_peak,taraskin97ab,wischnewski97,dellanna97}.
Various theoretical approaches have been proposed to explain this peak,
such as localized vibrational modes or scattering of acoustic modes,
but so far no clear picture has emerged yet.  Recently also computer
simulations have been used in order to gain insight into the mechanism
that gives rise to this peak, but due to the high cooling rates with
which the samples were prepared (on the order of $10^{12}$K/s) and
small system sizes (20-40\AA) the results of these investigations were
not able to give a final answer either~\cite{taraskin97ab,dellanna97}.
In the present work we present the results of a large scale computer
simulation of supercooled silica. At the temperature investigated we
are still able to fully equilibrate the sample, thus avoiding the
problem of the high cooling rates, and by using a large system we can
minimize the possibility of finite size effects~\cite{horbach96}. Thus,
by making a large computational effort, we are able to study the
dynamics of this strong glassformer in a frequency and wave-vector
range which is not accessible to real experiments and can therefore
investigate the properties of the boson-peak in greater detail than was
possible so far.

\section{Details of the Simulations}

The silica model we use for our simulation is the one proposed by van
Beest {\it et al.}~\cite{beest90} and has been shown to give a good
description of the static properties of silica glass~\cite{vollmayr96}
as well as the dynamical properties of the supercooled melt, such as the
activation energy of the diffusion constant~\cite{horbach97}.
In this model the potential between ions $i$ and $j$ is given by 
\begin{equation}
\phi(r_{ij})=\frac{q_i q_j e^2}{r_{ij}}+A_{ij}e^{-B_{ij}r_{ij}}-
\frac{C_{ij}}{r_{ij}^6}\quad .
\label{eq1}
\end{equation}
The values of the parameters $q_i$, $A_{ij}$, $B_{ij}$ and $C_{ij}$ can
be found in the original publication~\cite{beest90}. The non-Coulombic
part of the potential was truncated and shifted at 5.5\AA. The
simulations were done at constant volume and the density of the system
was fixed to 2.3 g/cm$^3$. The system size was 8016 ions, giving a size
of the box of (48.37\AA)$^3$, and the equations of motion were
integrated over $4\cdot 10^6$ times steps of 1.6fs, thus over a time
span of 6.4ns. This time is sufficiently long to fully equilibrate the
system at 2900K, the temperature considered in this
study~\cite{horbach97}. More details on the simulations can be found in
Ref.~\cite{horbach98}.

\section{Results}

In the present work we study the dynamics of the system by means of
$J_L(q,\nu)$ and $J_T(q,\nu)$, the longitudinal and transverse
current-current correlation functions for wave-vector $q$ at frequency
$\nu$~\cite{boon80}. These are defined as the longitudinal and
transverse part of the current-current correlation function, i.e.
\begin{equation}
J_{\alpha}(q,\nu)=N^{-1}\int_{-\infty}^{\infty} dt \exp(i2\pi\nu t) 
\sum_{kl} \langle {\bf u}_k(t) \cdot {\bf u}_l(0) \exp(i{\bf q} \cdot
[{\bf r}_k(t)-{\bf r}_l(0)]) \rangle \quad,
\end{equation}
where ${\bf u}_k(t)$ is equal to ${\bf q}\cdot \dot{\bf{r}}_k(t)/q$ for
$\alpha=L$ and equal to ${\bf q}\times \dot{{\bf r}}_k(t)/q$ for
$\alpha=T$. Note that $J_L(q,\nu)$ is also equal to $\nu^2 S(q,\nu)/q^2$,
where $S(q,\nu)$ is the dynamical structure factor as
measured in scattering experiments. In the following we will focus on
the silicon-silicon correlation only, but we have found that the
oxygen-oxygen correlation function behave very similarly.

In Fig.~1 we show the frequency dependence of $J_L(q,\nu)$ for
wave-vectors between 0.13\AA$^{-1}$, the smallest wave-vector
compatible with our box, and 0.8\AA$^{-1}$, a wave-vector which is
still significantly smaller than the location of the first sharp
diffraction peak in $S(q)$, which is around 1.6\AA$^{-1}$. From the
figure we recognize that this correlation function has a peak at a
frequency $\nu_L(q)$ which increases with increasing $q$ and which
correspond to the longitudinal acoustic modes. A similar picture is
obtained for $J_T(q,\nu)$~\cite{horbach98}. The fact that also this
correlation function shows an acoustic mode shows that even at this
relatively high temperatures the system is able to sustain transverse
acoustic modes and thus is visco-elastic.

The boson-peak is seen best in the dynamic structure factor $S(q,\nu)$,
which is shown in Fig.~2 for small values of $q$. From this figure we
see that $\nu_{BP}(q)$, the location of the boson-peak, is $q$-dependent
in that it moves from small frequencies for small values of $q$ (dashed
lines) to a maximum frequency for $q\approx 0.39$\AA$^{-1}$ (bold
dotted line) and then back to small frequencies for large values of $q$
(solid lines) (see also inset of Fig.~3). Also included is the
location of the boson-peak as determined from neutron scattering
experiments at 1673K which is around 1.5~THz~\cite{wischnewski97}. We
will comment more on this work below.

From the wave-vector dependence of $\nu_L$ and $\nu_T$ we obtain the
dispersion relation which is presented in Fig.~3. We see that, as
expected, for small wave-vectors $\nu_L$ depends linearly on $q$. Also
included in the figure is a line with slope $c_L$=6370m/s, the
experimental value of the longitudinal sound velocity of silica at
around 1600K\cite{wischnewski97}. We see that the data points for
$\nu_L$ for small $q$ are very close to this line and thus we conclude
that the sound velocity of this system is very close to the one of real
silica, thus giving further support for the validity of the model
potential.

For the transverse acoustic modes the agreement between the experiment
and the simulation data is a bit inferior, but still good.  We also
note that for wave-vectors larger than 1.4\AA$^{-1}$, a bit less than
the location of the first sharp diffraction peak, $\nu_L$ and $\nu_T$
do not increase anymore. The reason for this is likely the fact that at
this $q$ value the system has a quasi-Brillouin
zone~\cite{taraskin97ab}.

Also included in the figure is $\nu_{BP}$, the location of the
boson-peak. We find that for large wave-vectors $\nu_{BP}$ is around
1.8~THz, a value that is a bit larger than the experimental value of
1.5~THz reported by Wischnewski {\it et al.} at
1673K~\cite{wischnewski97}. However, these authors also found that
$\nu_{BP}$ increases with increasing temperature and a rough
extrapolation of their data for $\nu_{BP}$ to T=2900K shows that a
value of 1.8~THz is quite reasonable, thus giving further support for
the validity of our model.

In the inset of Fig.~3 we show the dispersion curves at small values of
$q$. Interestingly we observe that $\nu_{BP}(q)$ shows a maximum at around
$q\approx 0.39$\AA$^{-1}$. This maximum might be due to the fact that
the mechanism leading to the boson-peak is particularly effective at
this wave-vector or that there are {\it two} mechanisms giving rise to
the boson-peak, one dominating at small wave-vectors and the second one
dominating at larger wave-vectors and that in the vicinity of $q\approx
0.39$\AA$^{-1}$ the sum of the contribution of the two mechanisms is
largest.

Finally we mention that at small wave-vectors the curve for $\nu_{BP}(q)$
seems to join smoothly the one for $\nu_L$. Within the accuracy of our
data it is not clear, whether the boson-peak and the longitudinal
acoustic mode become identical or whether the boson-peak ceases to
exist for wave-vectors smaller than approximately 0.2\AA$^{-1}$ and
thus we cannot use this feature to exclude one of the possible
mechanisms proposed for the boson-peak.

To summarize we can say that our simulation allows to investigate the
dynamics of supercooled silica in a wave-vector and frequency range
which is not accessible to real experiments. We find that the
dispersion relations for the longitudinal and transverse acoustic modes
agree very well with the experimental values and that $\nu_{BP}$
shows a maximum at around 0.39\AA$^{-1}$. This $q$ corresponds to a
length scale on the order of 15\AA.  Thus we have evidence that the
mechanism leading to the boson-peak is very effective on the length
scale of several tetrahedra.

Acknowledgements: We thank U. Buchenau, G. Ruocco and F. Sciortino for
valuable discussions and the DFG, through SFB 262, and the BMBF,
through grant 03N8008C, for financial support.

\newpage
%\section*{Figures}
\begin{figure}[f]
\vspace{.3cm}
\psfig{file=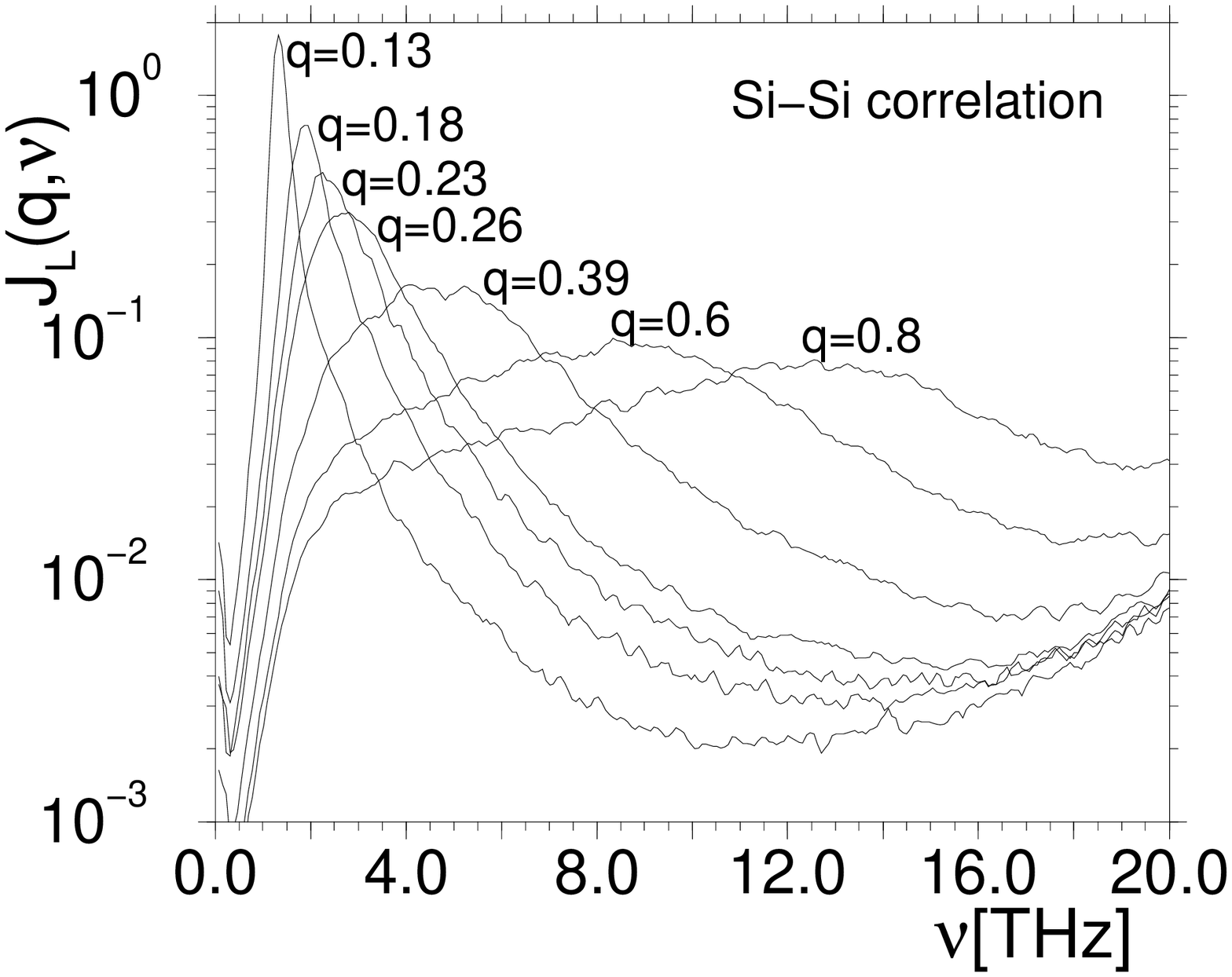,width=14cm,height=10.0cm}
       \caption{Frequency dependence of the longitudinal
current-current correlation for different wave-vectors $q$.}
\label{fig1}
\end{figure}

\begin{figure}[f]
\vspace{-0.0cm}
\psfig{file=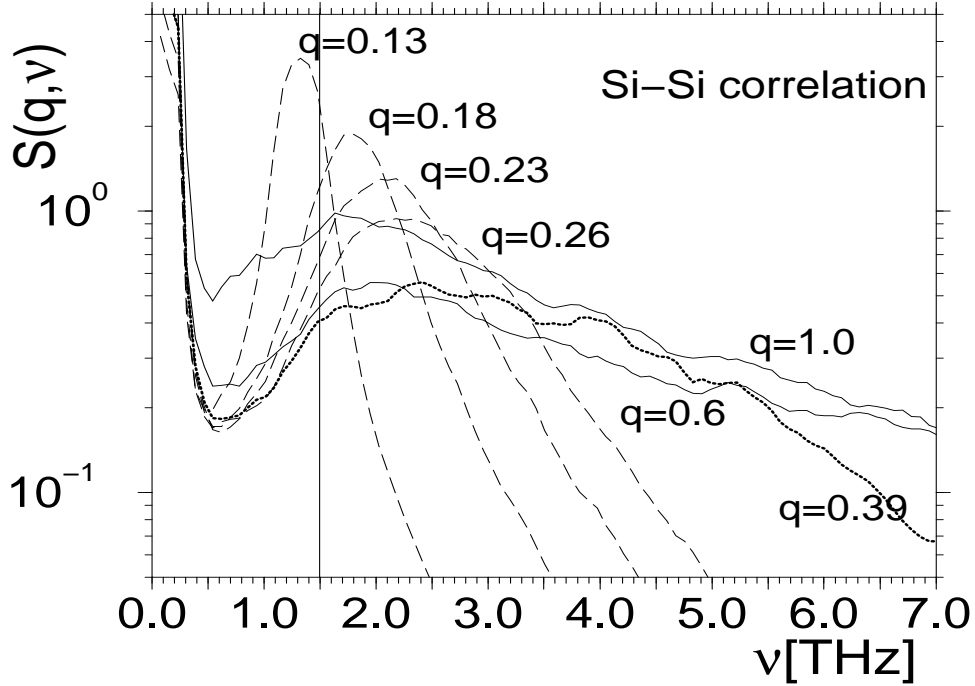,width=14cm,height=10.0cm}
\caption{Frequency dependence of the dynamic structure factor for
different wave-vectors $q$. The vertical line is the location of the
boson-peak at 1673K as determined from the neutron scattering
experiments~[3].}
\label{fig2}
\end{figure}

\begin{figure}[f]
\vspace{.03cm}
\psfig{file=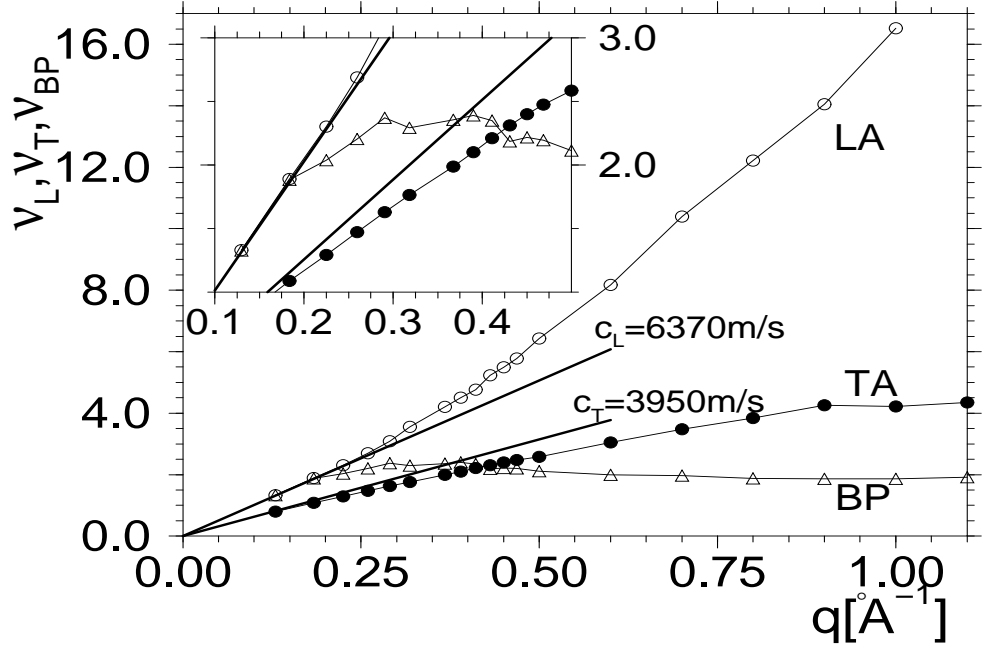,width=14cm,height=10.0cm}

\caption{Wave-vector dependence of $\nu_L$ (open
circles), $\nu_T$ (filled circles) and $\nu_{BP}$ (open triangles).
The bold solid lines are the dispersion relations for the longitudinal
and transverse acoustic modes (Ref.~[3]). Inset:
enlargement of the curves at small $q$.}
\label{fig3}
\end{figure}

\end{document}